\documentclass[11pt]{amsart}
\usepackage{amsfonts,amssymb,amsmath}

\newtheorem{theorem}{\bf Theorem}[section]

\newtheorem{remark}{\bf Remark}[section]
\newtheorem{lemma}{\bf Lemma}[section]
\newtheorem{definitions}{\bf Definition}[section]
\numberwithin{equation}{section}

\newcommand{\R}{\mathbb{R}}

\title[Portfolio optimization with fixed income securities]
{ A Risk-Sensitive Portfolio Optimization Problem with Fixed
Incomes Securities }

{\author{Mayank Goel and K. Suresh Kumar}
\address{ Research Analyst, Quantitative Strategies Group,
BA Continuum Solutions
Pvt. Ltd.(A non bank subsidiary of  Bank of America),
Mumbai-400072}
\address{Department of
Mathematics, Indian Institute of Technology Bombay, Mumbai -
400076, India. }}

\email{mayank.goel@bankofamerica.com, suresh@math.iiitb.ac.in
(corresponding author)}
\date{}

\begin{document}

\maketitle

{\bf Abstract.} We discuss a class of risk-sensitive portfolio
optimization problems. We consider the portfolio optimization
model investigated by Nagai in 2003. The model by its nature can
include fixed income securities as well in the portfolio. Under
fairly general conditions, we prove the existence of optimal
portfolio in both finite and infinite horizon problems.\par {\bf
Key words.} Risk-sensitive control, fixed income securities, non
stationary optimal strategies.\par {\bf AMS 2000 subject
classification.} 91B28, 93E20, 49L20, 35K55, 60H30.
\section{Introduction}

In this paper, we study a class of portfolio optimization problems
in continuous trading framework where the returns of the
individual assets are explicitly being affected by underlying
economic factors. The continuous time portfolio management has its
origin in the pioneering work of Merton, see \cite{m1,m2}. Since
then there were several contributions to the stochastic control
applications to portfolio management, see \cite{kara2,korn2} for
details. But most of these works deal with equities. Literature on
portfolio optimization with fixed income assets is limited. A
stochastic control model suitable for fixed income assets was
first formulated by Merton \cite{m1}.  Bielecki and Pliska in
\cite{bp1} and later in \cite{bp2}, investigated the following
linear version of Merton's model \cite{m1} with risk-sensitive
criterion,
\begin{equation*}\hspace{-.1in}\left\{ \begin{array}{lll}
\displaystyle{\frac{dS_i(t)}{S_i(t)}}&=&\displaystyle{
(a+AX(t))_{i}dt+\sum_{k=1}^{m+n}\sigma_{ik}dW_{k}(t),~~~~~S_{i}(0)=s_i,~~i=1,2,\cdots,m,}\\
\displaystyle{  dX(t)}&=&\displaystyle{(b+BX(t))dt+\Lambda
dW(t),~~~~X(0)=x,} \\
& & X(\cdot) =(X_1(\cdot),X_2(\cdot),\cdots,X_n(\cdot)),
\end{array}\right. \end{equation*}
where $S_i(t)$ denote the price of $ith$ security and $X_j(t)$ the
level of the $jth$ factor at time $t$ and $W(\cdot)$ is an
$\R^{m+n}$- valued standard Brownian motion with components
$W_k(\cdot)$. In \cite{bp2}, authors improved their earlier work
\cite{bp1} by relaxing the assumption $\Sigma \, \Lambda^{\perp}
\, = \, 0$. Hence, the portfolio model become capable of
incorporating fixed income securities such as rolling horizon
bonds ( it is a portfolio of bonds). \par

Also Nagai in \cite{nagai2}, considered the following  general
diffusion model and addressed the portfolio optimization problem
with risk-sensitive criterion. They assumed that the set of
securities includes one bond, whose price is defined by the ODE:
\begin{equation*}
dS_{0}(t)=r(X(t))S_{0}(t)dt,~~~S_{0}(0)=s_{0},
\end{equation*}
where $r(\cdot)$ is a nonnegative bounded function. The other
security prices $S_i(\cdot),~~i=1,2,\cdots,m$ and factors
$X(\cdot)$ are assumed to satisfy the SDEs
\[ \left\{
\begin{array}{lll}
dS_i(t)&=&S_{i}(t)[g_{i}(X(t))dt+\sum_{k=1}^{m+n}\sigma_{ik}(X(t))dW_{k}(t)],\\
S_i(0)&=&s_i,~~i=1,2,\cdots, m,\\
dX(t)&=&b(X(t))dt+\lambda(X(t))dW(t),\\
X(0)&=&x\in \R^{n}.
\end{array}
\right. \]
 Nagai proved the existence of optimal portfolios under
the following assumptions:
\begin{itemize}
\item[ (i) ] The functions $g, \, \sigma, \, b, \, \lambda$ are
Lipschitz continuous and $\sigma \sigma^{\perp}, \ \lambda
\lambda^{\perp}$ are uniformly elliptic. \item[ (ii) ] There
exists $r_0$ and $\kappa$ positive such that
\[
\frac{1}{2} \ {\rm tr}( \lambda  \lambda^{\perp} (x)) \, + \,
x^{\perp}  [b(x)  -  \lambda  \sigma^{\perp} (\sigma
\sigma^{\perp})^{-1} (g \, - \, r \, \bar{1})(x)] \, + \,
\frac{\kappa}{2}  \frac{x^{\perp} \lambda  \lambda^{\perp} (x)
x}{\sqrt{ 1+ \|x\|^2}} \ \leq \ 0
\]
for all $\|x\| \geq r_0$, $\bar{1} \, = \, ( \ \overbrace{1,
\cdots, 1}^m \ )^{\perp}$ \item[ (iii) ] Let $\hat{u}$ is the
solution to (\ref{s3pde2}), then
\[
\frac{4}{\theta^2} ( g- r \bar{1})^{\perp} (\sigma
\sigma^{\perp})^{-1}  (g - r \bar{1}) \ - \ (\nabla
\hat{u})^{\perp}  \lambda  \sigma^{\perp}  (\sigma
\sigma^{\perp})^{-1} \sigma \lambda^{\perp} \nabla \hat{u} \,
\rightarrow \infty \ {\rm as}\ \|x\| \rightarrow \infty \, .
\]

\end{itemize}

 Ergodic risk sensitive control problem for
the linear case is well studied, see \cite{bp1,bp2, fleming4,
fleming2} for example. But for the nonlinear case, most of the
related works deals with the small noise case, see for example
\cite{fleming4,fleming3,paul}. The nonlinear case, suited for the
continuous portfolio optimization is studied in \cite{flemingmc}
and later in \cite{nagai2}. The work \cite{flemingmc} also assumes
the a condition which is similar to the condition (ii) in
\cite{nagai2} given above. In this paper we consider the model
given in \cite{nagai2}. Our main contribution is that we prove the
existence of ergodic optimal investment strategy without the
assumption (ii) and the assumption (iii) replaced with the
assumption (A3) which is standard in the literature of stochastic
control.

Rest of our paper is organized as follows: In Section 2, we give a
formal description of the problem. In Section 3, we investigate
the finite horizon problem. We prove the existence of optimal
investment strategy in Theorem 3.1 and give an explicit form for
the optimal investment strategy in Theorem 3.2. In Section 4, we
prove the existence of optimal non stationary investment strategy
under (A1)-(A3). Note that the main challenge is in establishing
the uniqueness of the pde (\ref{s3pde2}). This is achieved without
the condition (ii) of \cite{nagai2} in Theorem 4.1.
\section{Problem Formulation}

\noindent  We consider an economy with $m \geq 2 $ securities and $ n\geq
1$ factors, which are continuously traded on a frictionless market.
All traders are assumed to be price takers. The set of securities may include
stock, bonds and savings account and the set of factors may include dividend
yields, price-earning ratios, short term interest rates, the rate of
inflation.\par
Let $S_{i}(t)$ denote the price of $ i$th security
and $X_{j}(t)$, the level of the $j$th factor at time t.
Dynamics of the security prices and factors are assumed to follow SDE
given by\\
\begin{equation} \label{s2sde1}
\left\{
\begin{array}{lll}
dS_0(t) & = & r(X(t)) \, d t, \ S_0(0) \, = \, s_0 > 0 \, , \\
 \displaystyle{\frac{dS_{i}(t)}{S_{i}(t)}}&=&
 \displaystyle{{a}_{i}(X(t))dt +
\sum_{k=1}^{m+n}{\sigma}_{ik}(X(t))dW_{k}(t),}\\
& & S_{i}(0)=s_{i} > 0,\;\;\;
i=1,2,\cdots,m,\\
d{X}_{i}(t)&=& \displaystyle{\mu_{i}(X(t)) dt +
\sum_{k=1}^{m+n}{\lambda}_{ik}(X(t)) dW_{k}(t),} \\
& &  X_{i}(0)=x_{i} , \;\;\; i=1,2,\cdots,n,
\end{array}
\right.
\end{equation}
\noindent where $a=(a_1,\cdots,a_m)^{\perp},~~\mu=(\mu_1,\cdots,
\mu_n)^{\perp},~~\sigma =[\sigma_{ij}]$ and
$\Lambda=[\lambda_{ij}]$ with $a: \R^n \rightarrow \R^m,~~\mu:
\R^n \rightarrow \R^n,~~\sigma :\R^n \rightarrow \R^{m\times
(m+n)}$, $\Lambda : \R^n \rightarrow \R^{n\times (m+n)}$ and $r :
{I\!\! R}^n \to \R$.

\noindent We assume that \\
{\bf (A1)} The functions $a_i, \ \mu_i,\sigma_{ij}, \lambda_{ij}$
are bounded  Lipschitz  continuous and  $r$ is  positive bounded
measurable.

\noindent {\bf (A2)} The functions $\sigma \sigma^{\perp}  , \
\Lambda {\Lambda}^{\perp}$ are  uniformly elliptic with uniform
ellipticity constant $\delta_0 > 0$.

\noindent Under (A1) and (A2), the SDE (\ref{s2sde1}) has  unique
strong solution.

If $n_i(t)$ denote the amount held by the investor in the $ith$ security
at time $t$, then the wealth $V(t)$ of the investor at time $t$ is
given by
\begin{equation*}
V(t)=\sum_{i=0}^{m}n_{i}(t)S_{i}(t).
\end{equation*}
Set $h_i(t)=\frac{n_i(t)S_i(t)}{V(t)},$ i.e., $h_i(t)$ is the fraction
of the wealth in the $ith$  security at time $t$. Then for a self
financing strategy wealth equation takes the form
\begin{equation}\label{s2sde2}
\left\{
\begin{array}{lll}
dV(t) & = & V(t)\, [ r(X(t)) \, + \, h(t)^{\perp}(a(X(t)) \, -
\, r(X(t)) \, \bar{1} ] \, dt \\
&&\\
 & & + \ V(t) \, h(t)^{\perp} \, \sigma(X(t)) \, d W(t) , \
V(0) = v > 0 \, ,
\end{array}
\right.
\end{equation}
where  $h(t) \ = \ (h_1(t), \cdots , h_m(t))^{\perp}$.

We use the following admissibility conditions for the investment
process $h(\cdot)$.\par
\begin{definitions} An investment process $h(\cdot)$ is
{\bf admissible } if the following conditions are satisfied:
\begin{itemize}
\item[(i)]  $h(\cdot)$ takes values in  $\R^m $. \item[(ii)] The
process $h(\cdot)$ is
progressively measurable with respect to the filtration\\
${\mathcal{G}}_t =\sigma(S_1(s),\cdots,S_m(s),X(s)|s\leq t).$
\item[(iii)]
$E\left(\int_{0}^{T}||h(s)||^{2}ds\right)<\infty,~~~\forall~ T$.
\end{itemize}
\end{definitions}
\noindent The class of admissible investment strategies is denoted
by ${\mathcal{H}}$. \par
For a prescribed admissible strategy $h(\cdot)$ (see
[\cite{fleming4} p.162] for the definition of prescribed strategy)
there exists a unique strong and almost surely positive solution
$V(\cdot)$ to the SDE (\ref{s2sde2})  see, [\cite{strook} p.192] .
Also for $h (\cdot) \in {\mathcal{H}}$, the SDE (\ref{s2sde2})
admits a unique weak solution.  For an admissible strategy
$h(\cdot)$ and for the initial conditions $x\in \R^n$ and $v>0$,
the risk-sensitive criterion for the horizon $[0,T]$ is given by
\begin{equation}\label{fpay}
{J_{\theta}}^{T}(v,x,h(\cdot))=\left(\frac{-2}{\theta}\right) \ln
E^{h(\cdot)}[e^{-(\frac{\theta}{2})\ln V(T)}|V(0)=v, X(0)=x].
\end{equation}
For the infinite horizon problem, the criterion is
\begin{equation}\label{infpay}
J_{\theta}(v,x,h(\cdot))=\liminf_{T \rightarrow \infty }\left(
\frac{-2}{\theta}\right)T^{-1} \ln
E^{h(\cdot)}[e^{-(\frac{\theta}{2})\ln V(T)}|V(0)=v, X(0)=x].
\end{equation}
\noindent We assume that $\theta>0$, i.e., the investor is risk
averse. Now the investor's optimization problem, is as follows:\\
\noindent For finite horizon
\begin{equation*}
\max_{h(\cdot)\in {\mathcal{H}}}J_{\theta}^{T}(v,x,h(\cdot))
\end{equation*}
\hspace{2.25in}subject to (\ref{s2sde1}) and (\ref{s2sde2}),\\
\noindent for infinite horizon
\begin{equation*}
\max_{h(\cdot)\in {\mathcal{H}}}J_{\theta}(v,x,h(\cdot))
\end{equation*}
\hspace{2.25in} subject to (\ref{s2sde1}) and (\ref{s2sde2}).
\begin{definitions}
 (i)An admissible strategy $h^{*}(\cdot)$ is said to be
optimal for the finite horizon problem if
$${J_{\theta}}^{T} (v,x,h(\cdot))\leq
{J_{\theta}}^{T}(v,x,h^{*}(\cdot)),~~~\forall \mbox{ admissible
}~~ h(\cdot).$$

(ii)An admissible strategy $h^{*}(\cdot)$ is said to be optimal
for the infinite horizon problem if
$$J_{\theta} (v,x,h(\cdot))\leq
J_{\theta}(v,x,h^{*}(\cdot)),~~~\forall \mbox{ admissible }~~
h(\cdot).$$

\end{definitions}

\section{ Finite Horizon Problem }
\setcounter{equation}{0}
In this section, we consider the finite horizon problem described in
the previous section. Our objective is to prove the existence of
optimal investment strategies for the payoff function
\begin{equation*}
J_{\theta}^T(v,x,h(\cdot))= \frac{-2}{\theta}\ln
E^{h(\cdot)}[e^{-(\frac{\theta}{2})\ln V(T)}|V(0)=v, X(0)=x].
\end{equation*}
The above optimal control problem is equivalent to minimize over
$h(\cdot) \in {\mathcal H}$, the  objective function
$${E}^{h(\cdot)}[{V(T)}^{\frac{-\theta}{2}}
|V(0)=v,X(0)=x],$$ \noindent where $(X(\cdot),V(\cdot))$ is
governed by (\ref{s2sde1}) and (\ref{s2sde2}).

We investigate the optimization problem by studying the corresponding
Hamilton Jacobi Bellman(HJB) equation given by\\
\begin{equation}\label{HJB}
 0 \ = \ \frac{\partial \phi}{\partial t} \ + \
 \inf_{h \in \R^m }{L}^{h(.)} \phi(t,x,v), ~~~\phi(T,x,v)={v}^{-(\theta / 2)}
 ~~\mbox{ for } t>0, x
 \in \R^{n}, v>0,
 \end{equation}
 where
\begin{eqnarray*}
L^{h}\phi &=&
 [r(x) \, + \, h^{\perp} (a(x) - r(x) \, \bar{1})] \, v  \frac{\partial\phi}{\partial
v} \, + \,  \sum_{i=1}^{n} \mu_i (x) \frac{\partial \phi}{\partial
x_i} \\
& &+ \frac{1}{2}h^{\perp} \sigma(x)\sigma(x)^{\perp} h
v^2\frac{{\partial}^{2}\phi}{\partial v^2}
+\frac{1}{2}\sum_{ij=1}^{n}m_{ij}(x)
 \frac{{\partial}^{2}\phi}{\partial x_i \partial x_j }\\
& & + \frac{v}{2}\sum_{i=1}^{n} \sum_{l=1}^{m}\sum_{k=1}^{m+n}\lambda_{ik}(x)\sigma_{lk}(x)
    h_l \frac{{\partial}^{2} \phi}{\partial x_i \partial v}, \\
m_{ij}(x) &= & \sum_{k=1}^{m+n}\lambda_{ik}(x)\lambda_{jk}(x).
\end{eqnarray*}
 We seek a solution to (\ref{HJB}) in the form \\
\begin{equation}\label{s2eq1}
\phi(t,x,v)= {v}^{-(\theta/2)}{e}^{-(\theta /2 )u(t,x)},
\end{equation}

\noindent for a suitable function $u$. Consider the following PDE
\begin{equation}
\left.
\begin{array}{lll} \label{s2pde1}
  0 & =& \displaystyle{ \frac{\partial u}{\partial t}} \ + \
\displaystyle{\sum_{i=1}^{n} \mu_i(x) \frac{\partial u}{\partial
x_i} +\frac{1}{2} \, \frac{-\theta}{2}\sum_{i,j=1}^{n}m_{ij}(x)
\frac{\partial u}{\partial x_i} \frac{\partial u}{\partial x_j} }\\
& &\\
& & \displaystyle{+ \, \frac{1}{2} \, \sum_{i,j=1}^{n}m_{ij}(x)
\frac{{\partial}^{2}u}{\partial x_i
\partial x_j} -K_{\theta}(x,\nabla u),
\;\; (t,x)\in (0, \infty)\times \R^{n},}\\
& &\\
 & &\displaystyle{u(T,x)=0, \;\;\;\; x \in \R^n,}\\
\end{array}
\right\}
\end{equation}
where,
\begin{equation}
\left.
\begin{array}{lll} \label{Kt}
\displaystyle{K_{\theta}(x,\nabla u)}&=&\displaystyle{\inf_{h \in
\R^m}\left[\frac{1}{2}\left(\frac{\theta}{2}+1\right)h^{\perp}
\sigma (x){\sigma (x)}^{\perp}h -h^{\perp}(a(x)- r(x) \bar{1}) - r(x)\right.}\\
& &\\
& &\displaystyle{\left.+ \frac{\theta}{4}\sum_{i=1}^{n}
\sum_{l=1}^{m} \sum_{k=1}^{m+n} \lambda_{ik}(x)\sigma_{lk}(x) h_l
\frac{\partial u}{\partial x_i} \right]}.
\end{array}
\right\}
\end{equation} \noindent Using straight forward
calculations, one can show that, the function $u\in
C^{1+\frac{\delta}{2},2+\delta}((0,T)\times\R^{n}), 0 < \delta < 1
$ is a solution to (\ref{s2pde1}) iff $\phi\in
C^{1+\frac{\delta}{2},2+\delta} ((0,T)\times\R^{n}) $ given by
(\ref{s2eq1}) is a solution to the HJB equation (\ref{HJB}).

\noindent  Set
$$u(t,x)=\frac{-2}{\theta} \ln
 \psi(t,x),~~(t,x) \in [0,\infty)\times \R^{n}.$$
\noindent Then we can show that $u\in
C^{1+\frac{\delta}{2},2+\delta}((0,T)\times \R^{n})$ is a solution
of (\ref{s2pde1}) iff $\psi \in C^{1+\frac{\delta}{2},2+\delta}
((0,T)\times\R^{n})$ is a positive solution of the PDE
\begin{equation}
\left.
\begin{array}{lll} \label{s2pde2}
0 &= & \displaystyle{\frac{\partial \psi}{\partial t} \, + \,
\frac{1}{2}\sum_{i,j=1}^{n} m_{ij}(x)
\frac{{\partial}^{2}\psi}{\partial x_{i}
\partial x_{j}}
+\sum_{i=1}^{n}\mu_i(x) \frac{\partial\psi}{\partial x_i}
 }\\
& &\\
& & + \, H(x, \psi , \nabla \psi),
\end{array}
\right\}
\end{equation}
where
\begin{equation}
\left.
\begin{array}{lll}\label{s2eq2}
\displaystyle{H(t,x,\psi ,\nabla
\psi)}&=&\displaystyle{\frac{\theta}{2} \inf_{h \in \R^m}
\left[\left\{\frac{1}{2}\left(\frac{\theta}{2} +1\right)h^{\perp}
\sigma(x) \sigma(x)^{\perp}h- h^{\perp} (a(x) - r(x) \bar{1}) \,
- \, r(x)\right\} \psi \right.}\\
& & \\
& & \displaystyle{\left.-h^{\perp} \sigma(x) \Lambda(x)^{\perp}
\nabla \psi \right]} .
\end{array}
\right\}
\end{equation}
\begin{lemma}\label{Lem1}
Assume (A1)-(A2). The PDE (\ref{s2pde2}) has unique solution $\psi
\in C^{1+\frac{\delta}{2},2+\delta}((0,T)\times \R^{n})$.
\end{lemma}
\noindent See [ \cite{ben}, pp. 94-97], [\cite{lady}, pp.419-423]
for a proof.

\begin{theorem}\label{The1}
Assume (A1)-(A2).  The  HJB equation (\ref{HJB}) has a unique
solution $\phi$ in $ C^{1,2}((0,T)\times \R^{n})$. Moreover
\begin{itemize}
\item[(i)] For $(s,x,v) \in [0, \ T) \times \R^n \times (0, \
\infty)$,
\[
\phi (s,x,v) \leq E^{h(.)} \left[ {V(T)}^{-(\theta/2)} |V(s)=v,
 X(s)=x \right]
\]
 for any admissible strategy $h(\cdot)$. \item[(ii)] If
$h^{*}(\cdot)$ is an admissible strategy such that
$$ L^{h^{*}} \phi(t,x,v) = \inf_{h \in \R^m} L^{h} \phi(t,x,v),~~~\forall~~t>0,~x\in
\R^n,~v>0 $$ then
 $\phi(s,x,v)= E^{h^{*}(\cdot)}\left[
{V^{*}(T)}^{-(\theta/2)}|V^{*}(s)=v,X(s)=x\right]$,\\
for any solution $V^{*}(\cdot)$ of (\ref{s2sde2}) corresponding to
$h^{*}(\cdot)$ and initial condition $(v,x)$.
\end{itemize}
\end{theorem}

\noindent {\bf Proof:} Existence of the solution of (\ref{HJB})
follows from Lemma \ref{Lem1}. Let $\phi\in
C^{1,2}((0,T)\times\R^n)$ be a solution to (\ref{HJB}). For each
admissible $h(\cdot)$ we have
$$0\leq  \frac{\partial \phi}{ \partial t} \ + \ L^{h(\cdot)}\phi(t,X(t),V(t)),~~t\geq 0 ,$$
where $(X(\cdot), V(\cdot))$ is given by (\ref{s2sde1})-
(\ref{s2sde2}) with initial conditions $X(s) = x, V(s) = v$.
\noindent For every integer $n \geq 1 $ define the stopping time
$${\tau}_{n}= T \bigwedge \inf\{ t \geq s~|~~~~||(X(t),V(t))||\geq n\},$$
where $||\cdot||$ is the usual norm in $\R^{n+1}$. Clearly,
${\tau}_{n} \uparrow T$.
Now using Ito's formula, we have  \\
$\phi({\tau}_{n},X({\tau}_{n}),V({\tau}_{n}))- \phi(s,x,v)$
\begin{eqnarray*}
=  \int_{s}^{T} \Big[ \frac{\partial \phi}{\partial t} \, + \,
L^{h(.)}\phi \Big] \, I_{[s,{\tau}_{n}]}(r)dr +
\int_{s}^{{\tau}_{n}}\left[ \sum_{i=1}^{n} {\lambda}_{i}(X(r))
\frac{\partial \phi}{\partial x_i} + h(r)^{\perp}\sigma(X(r))
V(r)\frac{\partial \phi}{\partial v} \right] dW(r)
\end{eqnarray*}
where $I_{[s,\tau_n]}$ denote the indicator function on
$[s,\tau_n]$ and $\lambda_i$ is the $ith$ row of matrix $\Lambda$.\\

Using $0 \leq \ \frac{\partial \phi}{\partial t} \, + \, L^{h(.)}
\phi(t,x,v),~~ \forall t>0,~v>0,~x\in\R^{n}$ and taking the
expectation on the both side, we have
$$E^{h(.)} [\phi (\tau_n, X(\tau_n),V(\tau_n))-
\phi(s,x,v)|V(s)=v,X(s)=x] \geq 0.$$ Now let $n \rightarrow \infty
$ we get,
\begin{eqnarray*}
0 & \leq & E^{h(.)} [ \phi (T, X(T),V(T))|V(s)=v,X(s)=x]\\
& & - E^{h(.)}[
\phi (s,x,v)|V(s)=v,X(s)=x].\\
&&\\
\phi (s,x,v) & \leq &  E^{h(.)} [ \phi (T,
X(T),V(T))|V(s)=v,X(s)=x].\\
&&\\
\phi (s,x,v) & \leq &  E^{h(.)}[ {V(T)}^{-(\theta /2)}
|V(s)=v,X(s)=x].\\
\end{eqnarray*}
For the proof of (ii), note that from the definition of $h^*(\cdot),$
we have
\begin{equation*}
L^{h^*(\cdot)}\phi(t,x,v)=0
\end{equation*}
Now using Ito's formula as above, it follows that
\begin{equation*}
\phi(s,x,v)=E^{h^*(\cdot)}\left[V^*(T)^{(-\theta/2)}|V^*(s)=v,
X(s)=x\right],
\end{equation*}
where $V^*(\cdot)$ is a solution to (\ref{s2sde2}) corresponding
to $h^*(\cdot)$. Hence
\begin{equation*}
\phi(s,x,v)=\inf_{h \in
\R^m}E^{h(\cdot)}\left[V(T)^{-(\theta/2)}|V(s)=v,~X(s)=x\right].
\end{equation*}
\hfill$\Box$\par

 \begin{theorem}\label{The2} Assume (A1)-(A2). Let $H_{\theta}(t, x)$ denote a
minimizing selector in (\ref{Kt}), that is,\\
\[
H_{\theta}(t, x)^{\perp} \ = \ \Big( \frac{2}{\theta +2} \Big)
\Big[ a(x) - r(x) \bar{1} \, + \, \frac{\theta}{2} \sigma
\Lambda^{\perp} \nabla u \Big] (\sigma \sigma^{\perp})^{-1}(x)\, .
\]
\noindent Then the investment process \\
\begin{equation}\label{s2eq3}
h_{\theta}(t):=H_{\theta}(t, X(t)),
\end{equation}
\noindent is optimal. i.e.\\
 \begin{equation}\label{Jt}
 {J_{\theta}}^{T}(v,x,h(\cdot)) \leq
 {J_{\theta}}^{T}(v,x,h_{\theta}(\cdot)),
 \end{equation}
 \noindent for all admissible  $h(\cdot), v>0, x\in
 \R^{n}$.
 \end{theorem}
{\bf Proof:}  Let $\phi$ be as in (\ref{s2eq1}) . Then it follows
from Theorem \ref{The1} that $\phi$ is the unique solution to the
HJB equation (\ref{HJB}). Since $H_{\theta}$ is a minimizing
selector in equation (\ref{Kt}), we have
$$L^{H_{\theta}(\cdot)}\phi(t,x,v)=\inf_{h \in \R^m} L^{h(\cdot)}\phi(t,x,v),~~\forall t>0,
~v>0,~x\in \R^{n}.$$ Now (i) and (ii) of Theorem \ref{The1}
implies that
\begin{equation*}
E^{h_{\theta}(\cdot)}\left[
{V^{*}(T)}^{-(\theta/2)}|V^{*}(s)=v,X(s)=x\right]
\leq E^{h(\cdot)} \left[ {V(T)}^{-(\theta/2)} |V(s)=v,
X(s)=x \right],
\end{equation*}
for all admissible  $h(\cdot)$ and $V^{*}(\cdot)$ is
    the unique solution to (\ref{s2sde2}) for the prescribed admissible
    strategy $h_{\theta}(\cdot)$. Hence, \\
\begin{equation*}
 {J_{\theta}}^{T}(v,x,h(\cdot)) \leq {J_{\theta}}^{T}(v,x,h_{\theta}(\cdot)),
 \end{equation*}
 \noindent for all admissible strategy $(h(\cdot), v>0, x\in \R^{n}$.
\hfill {$\Box$}
\section{ Infinite Horizon Problem }
\setcounter{equation}{0} In this section, we consider the infinite
horizon problem. The method is to treat the problem as the
asymptotic limit of the finite horizon problem. Thus we
investigate the asymptotic behavior of the HJB equation of the
finite horizon problem. Hence we require the following Lyapunov
type stability condition.

\noindent {\bf (A3)} There exists a function $v : \R^m \to \R$
such that\\
(i) $v \in C^2(\R^m), \ v \geq 0$ \\

\noindent (ii) The function $\|\nabla v \|$ has polynomial growth.

\noindent (iii) $L^{h, \omega} v (x) \to -\infty$ as $\|x\| \to
\infty$ for all $h$  and $\omega$, where
\begin{eqnarray*}
L^{h,\omega}\phi &= & \sum_{i=1}^n \Big[ \mu_i(x) \, + \,
\sum_{k=1}^{m+n} \lambda_{ik} (x) \omega_k  \, + \,
\frac{\theta}{2} \sum_{l=1}^n h_l \Big( \sum_{k=1}^{m+n}
\lambda_{ik}(x) \sigma_{lk}(x) \Big)  \Big]
\frac{\partial \phi }{\partial x_i} \\
{}&{}& ~~~~ + \, \frac{1}{2} \sum_{i=1}^n m_{ij}(x)
\frac{\partial^2 \phi}{\partial x_i
\partial x_j} \, .
\end{eqnarray*}

\noindent Consider the following auxiliary PDE\\
\begin{equation}
\left\{
\begin{array}{lll}\label{s3pde1}
\displaystyle{\frac{\partial \tilde{u}}{\partial t} }&=&
\displaystyle{\sum_{i=1}^{n} \mu_i(x) \frac{\partial
\tilde{u}}{\partial x_i} +\frac{1}{2}
\left[-\frac{\theta}{2}\sum_{i,j=1}^{n} m_{ij}(x) \frac{\partial
\tilde{u}}{\partial x_i}\frac{\partial \tilde{u}}{\partial x_j}
\right.}\\
& &\\
& & \displaystyle{\left.+ \sum_{i,j=1}^{n} m_{ij}(x)
\frac{{\partial}^{2} \tilde{u}}{\partial x_i
 \partial x_j}
\right]-K_{\theta} (x,\nabla \tilde{u}), \ t > 0, x \in
 \R^{n},}\\
 & &\\
& & \displaystyle{\tilde{u}(0, x)=0,\;\;\; \forall x \in \R^n,}
\end{array}
\right.
\end{equation}
We can show that $\tilde{u}\in C^{1+\frac{\delta}{2},2+\delta}((0,
T)\times \R^n)$ is a solution to (\ref{s3pde1}) iff $u\in
C^{1+\frac{\delta}{2},2+\delta}((0, T)\times \R^{n})$ is unique
solution to (\ref{s2pde2}). Hence (\ref{s3pde1}) has unique
solution $\tilde{u}\in C^{1+\frac{\delta}{2},2+\delta}((0,
T)\times \R^{n})$. Using Feynman-Kac representation of
(\ref{s3pde1}), see [\cite{kara1}, p.366] and (A3), we can show
that
 $\tilde{u}\geq 0,~~\frac{\partial \tilde{u}}{\partial t}\geq 0.$
Now we state the following estimate which is crucial to study the
asymptotic behavior of (\ref{s3pde1}).
\begin{lemma}\label{Lem2}
Let $\tilde{u}$ be the unique solution to (\ref{s3pde1}). Then for
each $c > 0$
\begin{equation*}
\begin{array}{ll}

\displaystyle{|\bigtriangledown_{x} \tilde{u}(t,x)|^{2} -
\frac{4(1+c)(\theta + 2)}{\theta \delta_{0} }\left|\frac{\partial
\tilde{u}(t,x)}{\partial t}\right| } & \\
~~\leq K \Big( |\nabla Q|^2_{2r} \, + \, | \nabla (\lambda
\lambda^{\perp})|^2_{2r} \, + \, | \nabla B|_{2r} \, + \,
|B|^2_{2r} + |U|_{2r} + \, |\nabla U|^2_{2r} +1 \Big), & \\
 ~~~~~~~~~~~~~~~~~~ t>0,~~ x \in B(0, r), &
\end{array}
\end{equation*}
where $\delta_{0}$ is the uniform ellipticity constant of $\Lambda
{\Lambda }^{\perp}$,
\[
Q (x) \ = \ \lambda \frac{\theta}{4} [ I \, - \,
\frac{\theta}{\theta +2} \sigma^{\perp} (\sigma
\sigma^{\perp})^{-1} \sigma ] \lambda^{\perp} \, ,
\]
\[
B (x)\ = \ \mu (x) \, - \, \frac{\theta}{\theta + 2} \lambda
\sigma^{\perp} ( \sigma \sigma^{\perp})^{-1}[ a(x) - r(x) \bar{1}]
\, ,
\]
\[
U(x) \ = \ \frac{1}{\theta + 2}( a - r \bar{1})^{\perp} (\sigma
\sigma^{\perp})^{-1} (a - r \bar{1}) + r(x) \,
\]
$| \cdot|_{2r} \ = \ \| \cdot \|_{L^{\infty}(B( 0, r))}$ and $K >
0$ is a constant depending on $ c, \, \delta_0 , n$.
\end{lemma}
\noindent The proof of Lemma 4.1 follows from the proof of [
\cite{nagai2}, Theorem 2.1 (i),  Remark (i)]. Now using the above
estimate we prove the following lemma, see appendix for the proof.

\begin{lemma}\label{Lem3} Let $\tilde{u}$  is be the solution to (\ref{s3pde1}) and
$x_0\in \R^{n}$, then there exists a subsequence $\{T_i\}\subset
\R_{+}$ such that $\tilde{u}(T_i,x)-\tilde{u}(T_i,x_0)$ converges
to a function $\hat{u}\in C^{2}(\R^{n})$ uniformly on compact sets
and strongly in $W_{loc}^{1,2}$ and $\frac{\partial
\tilde{u}(T_i,\cdot)}{\partial t}$ to $\rho \in \R$ uniformly on
each compact set. Moreover $(\hat{u}(\cdot),\rho)$ satisfies
\begin{equation}
\left.
\begin{array}{lll} \label{s3pde2}
\displaystyle{ \rho }&=& \displaystyle{\frac{1}{2}\sum_{i,j=1}^{n}
m_{ij}(x) \frac{{\partial}^{2} \hat{u}}{\partial x_i \partial x_j}
 -\frac{\theta}{4}\sum_{i,j=1}^{n} m_{ij}(x)\frac{\partial \hat{u}}{\partial
x_i}\frac{\partial \hat{u}}{\partial x_j} +\sum_{i=1}^{n}
\mu_{i}(x )\frac{\partial \hat{u}}{\partial x_i}
-K_{\theta}(x,\nabla \hat{u}),}\\
&&\\
&&\displaystyle{\lim_{||x||\rightarrow \infty}
\hat{u}(x)=\infty},~~x\in \R^n \, .
\end{array}
\right\}
\end{equation}
\end{lemma}

\noindent To show the uniqueness of the above PDE (\ref{s3pde2})
we rewrite (\ref{s3pde2}) as
\begin{eqnarray*}
\rho &=&\frac{1}{2}\sum_{i,j=1}^{n}
m_{ij}(x)\frac{{\partial}^{2}\hat{u}}{\partial x_i \partial x_j}
 -\inf_{\omega\in\R^{m+n}}\left[\frac{1}{\theta}||\omega||^{2}-\omega^{\perp}
 \Lambda(x)^{\perp}\nabla \hat{u}\right]
 +\mu(x)^{\perp} \nabla \hat{u}\\
 &&-\sup_{h \in \R^m}\left[ h^{\perp} (a(x) - r(x) \bar{1}) + r(x)-\frac{1}{2}\left(\frac{\theta}{2}
 +1\right)h^{\perp} \sigma (x) \sigma (x)^{\perp}h
-\frac{\theta}{2} h^{\perp} \sigma(x)\Lambda(x)^{\perp} \nabla \hat{u} \right] ,\\
&&\\
&&\displaystyle{\lim_{||x||\rightarrow \infty}
\hat{u}(x)=\infty,~~x\in\R^n.}
\end{eqnarray*}
Hence the PDE (\ref{s3pde2}) takes the form
\begin{equation}
\left.
\begin{array}{lll}\label{s3pde3}
\displaystyle{\rho} &=&\displaystyle{\frac{1}{2}\sum_{i,j=1}^{n}
m_{ij}(x)\frac{{\partial}^{2} \hat{u}}{\partial x_i \partial
x_j}}\\
 && \displaystyle{ +\sup_{\omega\in\R^{m+n}}\inf_{h \in \R^m}\left[
 \left( \mu(x)^{\perp} +  \omega^{\perp} \Lambda(x)^{\perp}
  + \frac{\theta}{2}h^{\perp}\sigma(x) \Lambda(x)^{\perp}
 \right)\nabla \hat{u}- \frac{1}{\theta}||\omega||^{2}\right.}\\
 &&\displaystyle{\left.+ \,
 \frac{1}{2}\left(\frac{\theta}{2} +1\right)h^{\perp} \sigma(x)
\sigma(x)^{\perp}h - h^{\perp}(a(x) - r(x) \bar{1}) - r(x) \right]}\\
&=&\displaystyle{\frac{1}{2}\sum_{i,j=1}^{n}
m_{ij}(x)\frac{{\partial}^{2} \hat{u}}{\partial x_i \partial
x_j}}\\
 && \displaystyle{+ \inf_{h \in \R^m} \sup_{\omega\in\R^{m+n}}\left[
 \left(\mu(x)^{\perp}+ \omega^{\perp} \Lambda(x)^{\perp} + \frac{\theta}{2}h^{\perp}
 \sigma (x) \Lambda (x)^{\perp}
 \right)\nabla \hat{u}- \frac{1}{\theta}||\omega||^{2}\right.}\\
 &&\displaystyle{\left. + \,
 \frac{1}{2}\left(\frac{\theta}{2} +1\right)h^{\perp} \sigma(x)
 \sigma (x)^{\perp} h - h^{\perp} (a(x) - r(x) \bar{1}) - r(x) \right],}\\
&&\\
&&\displaystyle{\lim_{||x||\rightarrow\infty} \hat{u}(x)=\infty .}
\end{array}
\right\}
\end{equation}
\noindent Consider the SDE
\begin{equation} \begin{array}{lll} \label{s3sde1}
dX_i(t)&=& \displaystyle{\left[ \mu_i(X(t))+ \sum_{k=1}^{m+n}
\lambda_{ik}(X(t))\omega_k(X(t))+ \frac{\theta}{2}
\sum_{l=1}^{m}\sum_{k=1}^{m+n} \lambda_{ik}(X(t))
\sigma_{lk}(X(t))h_l(t)\right]dt}\\
&&\\
&&+ \displaystyle{\sum_{k=1}^{m+n} \lambda_{ik}(X(t))dW_k(t), i
=1, \cdots, n.}
\end{array}
\end{equation}
Let ${\mathcal M}_1 $ denote the set of all Markov strategies in
$\mathcal{H}$ and
\[
{\mathcal M}_2 \ = \ \{ \omega : {I \! \! R} \to {I \! \! R}^{n+m}
\, | \ {\rm  measurable\ and} \  E \int^T_0 \| \omega(X(t)) \|^2
\, dt \ < \infty \ {\rm for\ all} \ T > 0 \} \, .
\]

\noindent For $h \in \R^m , \ w \in {I \! \! R}^{n+m}, \ \phi : {I
\! \! R}^n \to {I \! \! R}$, set
\begin{eqnarray*}
L^{h,\omega}\phi &= & \sum_{i=1}^n \Big[ \mu_i(x) \, + \,
\sum_{k=1}^{m+n} \lambda_{ik} (x) \omega_k  \, + \,
\frac{\theta}{2} \sum_{l=1}^n h_l \Big( \sum_{k=1}^{m+n}
\lambda_{ik}(x) \sigma_{lk}(x) \Big)  \Big]
\frac{\partial \phi }{\partial x_i} \\
{}&{}& ~~~~ + \, \frac{1}{2} \sum_{i=1}^n m_{ij}(x)
\frac{\partial^2 \phi}{\partial x_i
\partial x_j} \, .
\end{eqnarray*}
 and
\begin{equation*}
r(x, h,\omega)= \frac{1}{2}\left(\frac{\theta}{2}
+1\right)h^{\perp} \sigma(x) \sigma(x)^{\perp}h -
\frac{1}{\theta}||\omega||^{2}- h^{\perp} (a(x) - r(x) \bar{1}) -
r(x).
\end{equation*}


\noindent Let $\bar{\omega}(\cdot),~\bar{h}(\cdot)$ be such that
\[
\begin{array}{lll}
\displaystyle{ \sup_{\omega \in \R^{m+n}} \inf_{h \in \R^m} \Big[
L^{h, \omega } \hat{u} \, + \, r(h, \omega) \Big]} & = &
\displaystyle{ \inf_{h \in \R^m} \Big[
L^{h, \bar{\omega}(\cdot)} \hat{u} \, + \, r(h, \bar{\omega}(\cdot)) \Big]} \\
 \displaystyle { \ = \
\sup_{\omega \in \R^{m+n}} \Big[ L^{\bar{h}(\cdot) , \omega}
\hat{u} \, + \, r(\bar{h}(\cdot), \omega) \Big] } & = &
\displaystyle{ \inf_{h \in \R^m} \sup_{\omega\in \R^{m+n}} \Big[
L^{h, \omega} \hat{u} \, + \, r(h, \omega)\Big] \, } \\
& = & \displaystyle{L^{\bar{h}(\cdot) , \bar{\omega}(\cdot)}
\hat{u} \, + \, r(\bar{h}(\cdot), \bar{\omega}(\cdot))}
\end{array}
\]

\noindent  Fix $h(\cdot)\in {\mathcal M}_1$, let $X_1(\cdot)$
denote the process (\ref{s3sde1}) with initial condition $ x \in
\R^n $ corresponding to $(h(\cdot),\bar{\omega}(\cdot))$, then
using Ito's
formula, we have\\
$\hat{u}(X_1(T))-\hat{u}(x)$
\begin{eqnarray*}
&=&\int_0^{T}L^{h(\cdot),\bar{\omega}(\cdot)} \hat{u}(X_1(t))dt+
\mbox{Martingale (Zero-mean)}\\
&=& \int_0^{T}\left[L^{h(\cdot),\bar{\omega}(\cdot)}
\hat{u}(X_1(t))+r(X_1(t), h(X_1(t)) ,\bar{\omega}(X_1(t)))\right]dt\\
&&- \int_0^{T} r(X_1(t), h(X_1(t))
,\bar{\omega}(X_1(t)))dt+\mbox{Martingale (Zero-mean)}\\
&\geq& \inf_{h\in{\mathcal
M}_1}\int_0^{T}\left[L^{h(\cdot),\bar{\omega}(\cdot)}
\hat{u}(X_1(t))+r(X_1(t),h(X_1(t))
,\bar{\omega}(X_1(t)))\right]dt\\
&&- \int_0^{T} r(X_1(t),h(X_1(t))
,\bar{\omega}(X_1(t)))dt+\mbox{Martingale (Zero-mean)}\\
&=& \inf_{\omega\in{\mathcal M}_2}\sup_{h\in{\mathcal
M}_1}\int_0^{T} \left[L^{h(\cdot),\omega(\cdot)}
\hat{u}(X_1(t))+r(X_1(t),h(X_1(t))
,\omega(X_1(t)))\right]dt\\
&&- \int_0^{T} r(X_1(t),h(X_1(t))
,\bar{\omega}(X_1(t)))dt+\mbox{Martingale (Zero-mean)}\\
&=& T \rho -\int_0^{T} r(X_1(t),h(X_1(t))
,\bar{\omega}(X_1(t)))dt+\mbox{Martingale (Zero-mean) \, .}
\end{eqnarray*}
Taking expectation, we have
\begin{equation} \label{s3eq1}
E[\hat{u}(X_1(T)]- \hat{u}(x) \geq  \rho \, T -
E\left[\int_0^{T}r(X_1(t),h(X_1(t))
,\bar{\omega}(X_1(t)))dt\right] \, .
\end{equation}

\noindent Now mimicking the arguments in \cite{bo}(see appendix
for a proof), using (A3) we can show that $\hat{u} \in o(v(\cdot)$
and
\begin{equation} \label{s3eq2}
\lim_{T \rightarrow \infty} \frac{1}{T}E[\hat{u}(X_1(T))]= 0,
\end{equation}
Now divide (\ref{s3eq1}) by and let $T\rightarrow \infty$ we have
\begin{eqnarray*}
\rho \leq  \lim_{ T\rightarrow \infty} \frac{1}{T}E \left[\int_0^T
r(X_1(t),\bar{h}(X_1(t))
,\bar{\omega}(X_1(t)))\right]dt~~~\forall~~~h(\cdot)\in{\mathcal
M}_1.
\end{eqnarray*}
Therefore
\begin{eqnarray*}
\rho \leq \sup_{h(\cdot)\in{\mathcal M}_1}\underline{\lim}_{
T\rightarrow \infty} \frac{1}{T}E \left[\int_0^T
r(X_1(t),h(X_1(t)),\bar{\omega}(_1(t)))dt\right].
\end{eqnarray*}
Hence
\begin{equation}\label{4.6}
\rho\leq \inf_{\omega(\cdot)\in {\mathcal
M}_2}\sup_{h(\cdot)\in{\mathcal M}_1} \underline{\lim}_{
T\rightarrow \infty} \frac{1}{T}E \left[\int_0^T
r(X(t),h(X(t)),\omega(X(t)))dt\right],
\end{equation}
where $X(\cdot)$ is the process (\ref{s3sde1}) corresponding to
$(h(\cdot), \omega (\cdot))$. Now a similar argument shows that
\begin{equation}\label{4.7}
\rho\geq\sup_{h(\cdot)\in{\mathcal M}_1}\inf_{\omega(\cdot)\in
{\mathcal M}_2} \underline{\lim}_{ T\rightarrow \infty}
\frac{1}{T}E \left[\int_0^T r(X(t),h(X(t)),\omega(X(t)))dt\right].
\end{equation}
Combining (\ref{4.6}) and (\ref{4.7}), we get
\begin{eqnarray*}
\rho&=&\sup_{h(\cdot)\in{\mathcal M}_1}\inf_{\omega(\cdot)\in
{\mathcal M}_2} \underline{\lim}_{ T\rightarrow \infty}
\frac{1}{T}E
\left[\int_0^T r(X(t),h(X(t)),\omega(X(t)))dt\right]\\
&=&\inf_{\omega(\cdot)\in {\mathcal
M}_2}\sup_{h(\cdot)\in{\mathcal M}_1} \underline{\lim}_{
T\rightarrow \infty} \frac{1}{T}E \left[\int_0^T
r(X(t),h(X(t)),\omega(X(t)))dt\right]
\end{eqnarray*}
Let $( \rho^{'} , \psi)$ is another solution in the class $\R
\times C^2(\R_+^n) \cap o(\hat{u}(\cdot))$. Then using the similar
argument, one can easily check that
\begin{equation*}
\rho^{'} =\sup_{h(\cdot)\in{\mathcal H}}\inf_{\omega(\cdot)\in
{\mathcal M}} \underline{\lim}_{ T\rightarrow \infty} \frac{1}{T}E
\left[\int_0^T r(X(t),h(X(t)),\omega(X(t)))dt\right]=\rho
.\end{equation*} Let $h_1 \in {\mathcal M}_1$ be such that
\[
\rho \ = \ \inf_{ w \in \R^{n+m}} \Big[ L^{h_1(\cdot), w} \hat{u}
\, + \, r(x, h_1(x), w)\Big],
\]
 $w_1 \in {\mathcal M}_2$ be such that
\[
\rho \ = \ \sup_{ h \in \R^m} \Big[ L^{h, w_1(\cdot)} \psi \, + \,
r(x, h, w_1(x))\Big]
\]
and $X(\cdot)$ be the solution to (\ref{s3sde1}) corresponding to
$(h_1(\cdot), w_1(\cdot)$. Then
\begin{equation*}
L^{h(\cdot),\omega(\cdot)}(\hat{u}- \psi) \ \leq \ 0~~~\forall ~
h(\cdot)\in {\mathcal M}_1, ~~\omega(\cdot)  \in {\mathcal M}_2 \,
.
\end{equation*}
Thus $ \hat{u}(X(t))-\psi(X(t)),~t\geq 0$ is a submartingale
satisfying
\begin{equation*}
\sup_{t}E[|u(X(t))-\psi(X(t))|]\leq k\lim_{t\rightarrow
\infty}\frac{1}{t}\int_{0}^{t}||X(t)||^{2n}ds< \infty
,\end{equation*} for suitable $k>0,~n\geq1$. We use here the fact
that $\psi$ and $\hat{u}$  have polynomial growth. By the
submartingale convergence theorem, it converges a.s. Since
$\psi(x_0)= \hat{u}(x_0)=0$ and $X(\cdot)$ visits any arbitrarily
small neighborhood of zero infinitely often a.s., it can converge
only to zero. The same argument proves that $\psi- \tilde{u}$ is
identically zero: if not, $\psi- \hat{u} >\delta>0$ for some
$\delta$ in some open ball which is visited infinitely often
a.s.by $X(\cdot)$, contradicting the convergence of
$\psi(X(\cdot))-\hat{u}(X(\cdot))$ to zero. Hence   $\psi-
\tilde{u}$ is identically zero. Thus we have the following
theorem.
\begin{theorem}
Assume (A1)-(A3). The pde (\ref{s3pde1}) has a unique solution
$(\rho, \hat{u}) \in \R \times C(\R^n)$ satisfying $\hat{u}(x_0)
\, = \, 0$.
\end{theorem}
 \begin{theorem}\label{The3}
 Assume (A1)-(A3). Let $h_{\theta}(\cdot)$ be as in Theorem 3.2.
 Then:
 \begin{itemize}
 \item[(i).] For all $v>0$ and $x \in \R^{n}$ we have\\
\begin{eqnarray*}
{J_{\theta}}(v,x,h_{\theta}(\cdot))&=& \lim_{t \rightarrow
 \infty}\left(\frac{-2}{\theta}\right)t^{-1}\ln E^{h_{\theta}(\cdot)}\left[
 e^{-(\theta / 2)\ln V^*(t)}|V(0)=v, X(0)=x\right]\\
 &:=&\rho(\theta)
  \end{eqnarray*}
 where $V^*(\cdot)$ is the unique solution of (\ref{s2sde1}) corresponding to
 $h_{\theta}(\cdot)$ and the initial condition $(v,x).$
  \item[(ii).] The
admissible strategy $h_{\theta}(\cdot)$ is optimal.
\end{itemize}

\end{theorem}

\noindent {\bf Proof:} From Theorem 3.2, we have
\begin{equation} \label{s4eq1}
\frac{1}{T} J^T_{\theta}(x, v, h_{\theta}(\cdot)) \ \geq \
\frac{1}{T} J^T_{\theta}(x, v, h(\cdot))
\end{equation}
for all $h\cdot)$ admissible. Now using Theorem 4.1, we have
\begin{equation} \label{s4eq2}
\begin{array}{lll}
\frac{1}{T} J^T_{\theta}(x, v, h_{\theta}(\cdot)) & = &
\frac{1}{T} \frac{-2}{\theta} \ {\rm ln} \, \phi(T-t, x, v) \\
&=& \frac{1}{T} \ {\rm ln}\, v \, - \, \frac{1}{T} u(T-t, x) \\
& \to & \rho \ {\rm as}\ T \to \infty \, .
\end{array}
\end{equation}
Now from (\ref{s4eq1}) and (\ref{s4eq2}), we have
\[
\rho \ = \ \lim_{T \to \infty} \frac{1}{T} J^T_{\theta}(v, x,
h_{\theta}(\cdot)) \ \geq \ \liminf_{T \to \infty} \frac{1}{T}
J^T_{\theta}(v, x, h(\cdot))\, .
\]
Hence we have the theorem. \qed

\begin{remark}
We have shown that the optimal strategies in both finite horizon and
infinite horizon problems are functions of the economic factors only.
This happens since the economic factors are what which drives the asset
price movements. Another interesting observation is the same optimal
strategy works for both finite and infinite horizon problems.
\end{remark}

\begin{remark} (i) If we assume that $\sigma \Lambda^{\perp} \, \equiv \, 0 $, then strategy
given in Theorem 4.2 is stationary. But in this case portfolio
cannot include bonds.

\noindent (ii) Instead of  $\sigma \Lambda^{\perp} \, \equiv \, 0
$ if we assume the condition (ii) of \cite{nagai2}, then a close
mimicry of the proof of [ \cite{nagai2}, Theorem 4.1] we can show
that
\[
H_{\theta}(x) \ = \ \frac{\theta}{\theta + 2} (\sigma
\sigma^{\perp})^{-1} [ a(x) - r(x) \bar{1} - \frac{\theta}{2}
\sigma \lambda^{\perp} \nabla \hat{u}(x) \, .
\]
is an optimal stationary strategy.
\end{remark}

\section{Conclusion}
In this paper, we have investigated the risk-sensitive portfolio
optimization problem where the assets are explicitly depending on the
economic factors. Our portfolio model can also include fixed income
securities such as rolling horizon bonds. We prove the existence of
optimal investment strategies under very general conditions.\\

\section{Appendix}
\setcounter{section}{6}
 \setcounter{equation}{0}

\noindent {\bf Proof of Lemma 4.2.}\\

Set $\tilde{\phi}(T,x)=\tilde{u}(T,x)-\tilde{u}(T,x_0)$. Using
Lemma 4.1, it can be shown that $\{\tilde{\phi}(T, \cdot) | T > 0
\}$ is uniformly bounded and equicontinuous on compact subsets of
 $\R^n$. Therefore it has a subsequence
$\{\tilde{\phi}(T_i,\cdot)\}$ converging to a function
$\hat{u}(\cdot)\in C^{2}(\R^n)$ uniformly on each compact set.
Moreover, $\frac{\partial \tilde{u}}{\partial t}\geq 0 $ and by
Lemma \ref{Lem2} $\{\tilde{\phi}(T,\cdot)\}$ forms a bounded
subset of Hilbert space $W^{1,2}(B(0, R))$ for each $R > 0$ and we
see that there exists a subsequence (w.o.l.g
itself)$\{\tilde{u}(T_i ,\cdot)\}$ converging to $\bar{u} \in
W_{loc}^{1,2}(\R^n)$ weakly in $W_{loc}^{1,2}$ and strongly in
$L_{loc}^{2}$. Taking a further subsequence(w.l.o.g itself), we
can see that $\tilde{u}(T_i ,\cdot)\rightarrow \bar{u}(\cdot)$
a.s. and that $\bar{u} \equiv \hat{u}$.

Also we can show that $\nabla \tilde{\phi}(T_i ,\cdot)\rightarrow
\nabla \hat{u}(\cdot)$ strongly in $L_{loc}^{2}(\R^n)$.

\noindent Put $\xi(\cdot)=\frac{\partial \tilde{u}}{\partial t}$.
Then we obtain from (\ref{s3sde1})
\begin{equation*}
\begin{array}{lll}
\frac{\partial \xi}{\partial t}& = & \frac{1}{2}\sum_{ij=1}^{n}
m_{ij}(x)\frac{\partial^2 \xi}{\partial x_i
\partial x_j}+\sum_{i=1}^{n} \mu_{i}(x)\frac{\partial \xi}{\partial
x_i} -\frac{\theta}{2}\sum_{ij=1}^{n} m_{ij}(x) \frac{\partial
\tilde{u}}{\partial x_j}\frac{\partial \xi}{\partial x_i} \\
& &  -\frac{\theta}{2}\inf_{h \in \R^m}h^{\perp} \sigma (x)
\Lambda(x)^{\perp}\nabla \xi,
\end{array}
\end{equation*}
since $\xi$ is bounded on $(\epsilon,\infty)\times B(0, R)$
because of Lemma \ref{Lem2}, the regularity theorem for parabolic
equations implies that $\{\xi(T,\cdot)\}$ forms a family of Holder
equicontinuous functions on $(\epsilon,\infty)\times B(0, R)$ for
each $R$. Thus we have a subsequence (w.o.l.g. itself) $\{\xi(T_i
,\cdot)\}$ converging to a function $\rho\in C(\R^n)$ uniformly on
each compact set. Now take the limit along the subsequence in
\begin{equation}\left.\begin{array}{lll}\label{s6sde1}
\displaystyle{\frac{\partial \tilde{\phi}(T_i,x)}{\partial t} }&=&
\displaystyle{\sum_{i=1}^{n} \mu_{i} (x) \frac{\partial
\tilde{\phi}(T_i,x)}{\partial x_i} +\frac{1}{2}
\left[-\frac{\theta}{2}\sum_{i,j=1}^{n} m_{ij}(x) \frac{\partial
\tilde{\phi}(T_i,x)}{\partial x_i}\frac{\partial
\tilde{\phi}(T_i,x)}{\partial x_j}
\right.}\\
& &\\
& & \displaystyle{\left.+ \sum_{i,j=1}^{n} m_{ij}(x)
\frac{{\partial}^{2} \tilde{\phi}(T_i,x)}{\partial x_i
 \partial x_j} \right]- K_{\theta}
 (x,\nabla \tilde{\phi}),~~~~(T_i,x) \in (0, \infty
 )\times \R^{n},}\\
 & &\\
& & \displaystyle{\tilde{\phi}(0, x)=0,\;\;\; \forall x \in \R^n,}
\end{array}\right\}\end{equation}
we can see that $(\hat{u}(\cdot),\rho(\cdot))$ satisfies
(\ref{s3pde1}). Now we show that $\rho(\cdot)$ is a constant.\\
Fix $x^1 \in B(0, R_0)$. For $x\in B(0, R)$, for $R\geq R_0 $
\begin{eqnarray*}
\rho(x)&=& \lim_{n\rightarrow \infty}\frac{\partial
\tilde{\phi}(T_n,x)}{\partial
t}=\lim_{n\rightarrow\infty}\frac{\tilde{\phi}(T_n,x)}{T_n}\\
&=&
\lim_{n\rightarrow\infty}\frac{\tilde{\phi}(T_n,x)-\tilde{u}(T_n,x^1)}{T_n}
\  +  \  \lim_{n\rightarrow\infty}\frac{\tilde{\phi}(T_n,x^1)}{T_n}\\
&=& \lim_{n\rightarrow\infty}\frac{\nabla
\tilde{\phi}(T_n,x^1).(x-x^1)}{T_n} +\rho(x^1).
\end{eqnarray*}
Now from Lemma \ref{Lem2}, it follows that
\begin{equation*}
\lim_{n\rightarrow\infty}\frac{\nabla
\tilde{\phi}(T_n,x^1).(x-x^1)}{T_n}=0 ~~\mbox{ whenever }~~ x\in
B(0, R).
\end{equation*}
Therefore $\rho(x)=\rho(x^1)$ whenever $x \in B(0, R),$ for any $R\geq R_0.$\\
Since $R_0, \, R$ can be chosen arbitrary, we have
\begin{equation*}
\rho(x)=\rho(x^1)~~\forall~~y\in\R^n.
\end{equation*}
Hence $\rho$ is constant. \hfill $\Box$ \\

\noindent {\bf Proof of (\ref{s3eq2}).} From (A3), there exists $r
>0$ such that
\[
L^{h, \omega} v(x) \ \leq \ -1, \ {\rm whenever}\ \|x\| \geq r , h
\in \R^m , \ \omega \in \R^{n+m} \, .
\]
Let $X(\cdot)$ be the process corresponding to $(\bar{h}(\cdot),
\bar{\omega}(\cdot))$ with $X(0) \ = \ x , \ \|x\| \geq r$. Note
that
\[
\bar{h}(x) \ = \ \frac{2}{\theta +2} (\sigma \sigma^{\perp})^{-1}
[ a(x) - r(x) \bar{1} - \frac{\theta}{4} \Lambda \sigma^{\perp}
\nabla \hat{u}]
\]
and
\[
\bar{\omega}(x) \ = \ \frac{\theta}{2} \Lambda \nabla \hat{u}(x)
\, .
\]
From Lemma 4.1 and (A1), it follows that
\[
\| \nabla \hat{u}\|_{L^{\infty}(\R^n)} \ \leq \ c , \ {\rm for\
some} \ c
>0 \, .
\]
Hence there exits a constant $c_1 > 0$ such that
\[
\| \bar{h} \|_{L^{\infty}(\R^n)} \ + \ \| \bar{\omega}
\|_{L^{\infty}(\R^n)} \ \leq \ c_1 \, .
\]
Let $\tau_r$ be the first time the process $X(\cdot)$ hits the
ball $B(0, r)$.
 Using Ito's formula we have,
\[
E \hat{u}(X(\tau_r)) \ - \ \hat{u}(x) \ = \ - E \Big[
\int^{\tau_r}_0 r( X(s), \bar{h}(X(s)), \bar{\omega}(X(s))) \, ds
\]
Therefore from [ \cite{bo}, Lemma 4.1, p. 166], there exists
constants $c_2, \ c_3$ such that
\[
\hat{u}(x) \ \leq \ c_2 \, + \, c_3 \, v(x), \ \|x\| \geq r.
\]
i.e. $\hat{u} \in o( v(\cdot))$. Now mimicking the arguments from
[\cite{bo}, pp.165-168], the equation (\ref{s3eq2}) follows.

\bibliography{ref}
\bibliographystyle{plane}

\end{document}